# Transverse Spin Structure of the Nucleon

## Matthias Grosse Perdekamp[1] and Feng Yuan[2]

[1]Physics Department, University of Illinois at Urbana–Champaign, Urbana, Illinois 61801; email: mgp@illinois.edu

[2]Nuclear Science Division, Lawrence Berkeley National Laboratory, Berkeley, California 94720; email: fyuan@lbl.gov

## Abstract

We review the current status and future perspectives of theory and experiments of transverse spin phenomena in high-energy scattering processes off nucleon targets and related issues in nucleon structure and QCD. Systematic exploration of transverse spin effects requires measurements in polarized deep-inelastic scattering, polarized $pp$ collisions, and $e^+e^-$ annihilations. Sophisticated QCD-based techniques are also needed to analyze the experimental data sets.

## Contents

# 1. INTRODUCTION

Exploring the building blocks of nuclear matter, protons and neutrons, is of fundamental importance in science. Rutherford's (1) pioneering experiment 100 years ago revealed the structure of atomic matter, discovered the existence of small but massive nuclei at the core of atoms, and established the experimental scattering techniques employed over the following century to explore nuclear matter and its building blocks. Using probes of increasing energy, scattering experiments have revealed the quark and gluon structure of the nucleon in deep-inelastic scattering (DIS), the scattering of high-energy electron and muon beams off proton and neutron targets (2, 3). These experimental discoveries have led to the development of quantum chromodynamics (QCD) as the accepted theory describing the physics of strong interactions, such as the interactions between quarks and gluons (4–6).

The quark and gluon structure of the nucleon can be parameterized with the help of parton distribution functions, which describe the momentum distributions of partons inside the nucleon. A detailed quantitative understanding of these distributions has been gained through QCD-based analyses of a multitude of high-energy scattering experiments, including DIS, Drell–Yan lepton pair production, and inclusive processes (jet and electroweak boson production) in proton–proton ($pp$) collisions (7, 8).

In recent years, there has been significant progress in both experiment and theory toward extending the original Feynman parton picture to a full three-dimensional tomographic understanding of the partons inside the nucleon. These efforts represent the current forefront of the exploration of nuclear structure through scattering experiments a century after Rutherford's pioneering measurements. These extensions are necessary not only to reveal the internal structure of the nucleon in unprecedented depth and precision but also to explore ways of connecting the underlying parton distributions with nucleon properties. The latter are deeply connected to the associated strong interaction QCD dynamics. Assuming that the probed nucleon is moving in the $\hat{z}$ direction, the extension to the transverse direction can be either in the coordinate space or in the momentum space: Transverse momentum extensions to longitudinal momentum–dependent parton distribution functions are called transverse momentum–dependent (TMD) parton distributions (9–12), whereas transverse coordinate space extensions are referred to as generalized parton distributions (GPDs) (13–16).

Studying transverse nucleon spin structure provides a unique way to unveil the internal structure and parton correlations of the nucleon. Transverse spin, the spin component perpendicular to the nucleon momentum direction, is a natural vector to correlate with the transverse momentum of partons in TMD distributions or with their transverse position in GPDs. The first transverse spin measurements in high-energy hadron scattering were carried out in the late 1970s, when large single-spin asymmetries (SSAs) were observed in hadron production in $pp$ collisions, $p_{\uparrow}p \to h + X$ (17–22). An SSA is defined as an asymmetry between counting rates for a given reaction for either the probe or the target transversely polarized. In recent years, experimental and theoretical interest in transverse spin measurements has significantly increased. Experimentally, transverse spin studies have been extended to collider energies at the Relativistic Heavy Ion Collider (RHIC) at Brookhaven National Laboratory (BNL). Also at RHIC, large SSAs for charged and neutral hadrons have been observed in $pp$ scattering in the forward region of the polarized proton beam by the BRAHMS, PHENIX, and STAR experiments (23–29). In addition, high-energy lepton scattering experiments, including HERMES at DESY, COMPASS at CERN, and several experiments at Thomas Jefferson National Accelerator Facility (JLab), have observed large SSAs in semi-inclusive deep-inelastic scattering (SIDIS) of electrons or muons off transversely polarized nucleon targets (33–45). The observation of azimuthal correlations in back-to-back hadron pair production from $e^+e^-$ annihilation to dijets has also established significant transverse spin effects in the fragmentation of quarks with transverse spin to pions and kaons. These measurements became possible as a result of the high luminosity at the $B$ factories at KEK and SLAC and were carried out independently with the Belle and BaBar detectors (46–48). All these experimental results are intimately connected and have stimulated intense theoretical developments in the past few years. Today, 40 years after transverse spin phenomena in high-energy scattering were first observed, a QCD-based unified understanding of the broad range of transverse spin phenomena observed in SIDIS, $pp$, and $e^+e^-$ experiments seems to be emerging.

Transverse spin phenomena offer a unique opportunity to explore the QCD dynamics of quarks and gluons in hadron structure, to test advanced concepts of QCD factorization, and to investigate the universality of relevant parton distributions between different high-energy scattering processes. The associated transverse spin observables in high-energy scattering may well provide the only avenue to test certain unique QCD predictions. An important example is the expected sign change between SSAs observed in SIDIS and future measurements of SSAs in Drell–Yan lepton pair production in hadron collisions with transversely polarized proton targets or beams (49–51). This prediction is a direct consequence of the gauge invariance of QCD theory (51–54). It is also a manifestation of the underlying color dynamics in strong interaction theory, which often is hidden in spin-averaged cross sections. A second important example is the nontrivial fragmentation function that describes the correlation between the final-state hadron momentum and the fragmenting quark spin (55, 56). These correlations, confirmed by measurements from the Belle and BaBar experiments (46–48), may provide opportunities for developing a detailed understanding of the formation of hadrons from quarks and gluons in high-energy reactions. Future experimental exploration of transverse spin asymmetries with greatly improved statistical precision and enhanced instrumentation capabilities at JLab, Belle II, and a possible electron–ion collider (EIC) in China or the United States will lead to a deeper understanding and quantitative description of strong interactions in QCD and the quark and gluon substructure of the nucleon.

In this review, we attempt to describe the most important developments in recent years. We focus on the partonic structure of the transversely polarized nucleon. For additional information on other aspects of the three-dimensional structure of nucleons, we refer readers to recent reviews (57–65).

## 2. TRANSVERSE SPIN PHENOMENA IN EXPERIMENTS

Transverse spin phenomena observed in high-energy scattering represent the combined transverse spin–dependent mechanisms at work in the initial hard scattering reaction and the final-state fragmentation of quarks and gluons into experimentally observed hadrons. Phenomenological interpretation of transverse spin observables and model-independent separation of initial- and final-state transverse spin effects require the measurement of complementary observables in SIDIS, hadron–hadron scattering, and $e^+e^-$ annihilation, as well as their simultaneous QCD-based analysis. In the following sections, we introduce different experimental processes and use this overview to introduce relevant kinematics and nomenclature.

### 2.1. Lepton–Nucleon Scattering

The first class of experiments uses lepton scattering off nucleon targets, including elastic and inelastic scatterings, as shown in **Figure 1**. The use of elastic scattering to probe nucleon form factors has played a key role in revealing the finite size of the proton through the deviation of experimental observations from predictions for a pointlike particle (66, 67). Such experiments have provided much information on the internal nucleon structure and will continue to do so in the future via high-precision experimental data (68–71).

As the energy of the lepton beam is increased, lepton–nucleon scattering enters into the DIS regime (2, 3). In DIS, experiments measure structure functions, which can be interpreted as the momentum distribution of partons inside the nucleon, as Feynman had originally suggested. For

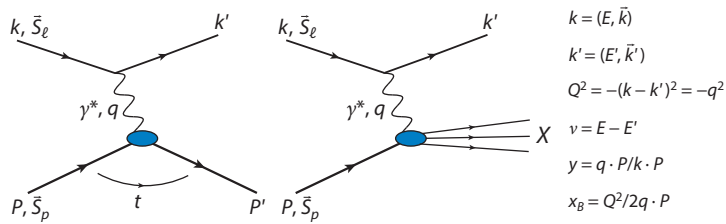

$$k = (E, \vec{k})$$
$$k' = (E', \vec{k}')$$
$$Q^2 = -(k - k')^2 = -q^2$$
$$\nu = E - E'$$
$$y = q \cdot P / k \cdot P$$
$$x_B = Q^2 / 2q \cdot P$$

**Figure 1**

Elastic scattering $\ell(k) + p(P) \rightarrow \ell(k') + p(P')$ and deep-inelastic scattering (DIS) off nucleon targets $\ell(k) + p(P) \rightarrow e(k') + X$ through virtual photon exchange. Elastic scattering probes the nucleon form factors, whereas DIS probes the Feynman parton distributions at the scale of $Q$ with parton momentum fraction $x_B$.

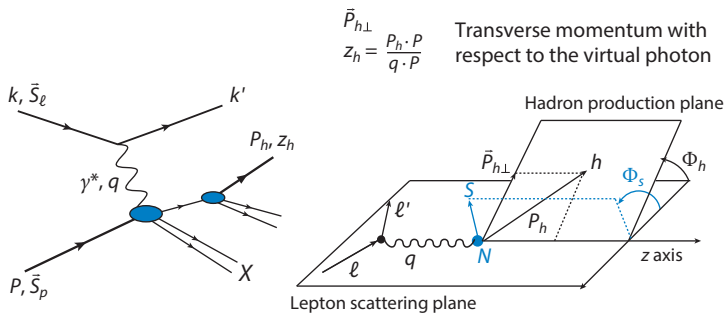

**Figure 2**

Semi-inclusive hadron production in deep-inelastic scattering off nucleon targets probes the transverse momentum–dependent (TMD) distributions, which also couple to the associated TMD fragmentation functions.

inclusive DIS processes, **Figure 1** summarizes the kinematic variables, among which Bjorken $x_B$ and $Q^2$ are the most important. $x_B$ is related to the momentum fraction of the nucleon carried by the parton, whereas $Q^2$ represents the scale of the parton distribution probed in this process.

In the past decade or so, with advances in both experiments and theory, two additional processes have been extensively studied in DIS experiments: (*a*) SIDIS (10–12) and (*b*) deeply virtual Compton scattering and deeply virtual meson production in exclusive DIS (60–63) (respective schematic diagrams are shown in **Figures 2** and **3**). In both processes, additional kinematic variables are measured. As a result, we are able to extend the one-dimensional Feynman parton distribution picture to a three-dimensional nucleon tomography. In particular, the transverse momentum of the final-state hadron $P_{b\perp}$ in SIDIS provides information on TMD distributions, whereas the recoil momentum $\Delta_T = P_T' - P_T$ in deeply virtual Compton scattering provides the coordinate space distributions of partons inside the nucleon in studies of GPDs. Below, we use both subscript $T$ and $\perp$ to represent the transverse index perpendicular to the beam direction. Both notations have been used in the literature, and they are interchangeable in this review.

GPDs and TMD distributions are intimately connected to each other and are unified under the concept of Wigner distributions. In the 1930s, Wigner (72) introduced the eponymous distributions to describe the phase-space distribution in quantum mechanics: $W(r, p) = \int d\eta e^{ip\eta} \psi^*(r - \frac{\eta}{2}) \psi(r + \frac{\eta}{2})$, where $r$ and $p$ represent the coordinate and momentum space variables, respectively, and $\psi$ is the wave function. When integrating over $r$ ($p$), one gets the momentum (probability) density from the wave function. Similarly, Wigner distributions may be defined for quarks and gluons (73–75): $W(x, k_\perp, r_\perp)$, where $x$ represents the longitudinal momentum

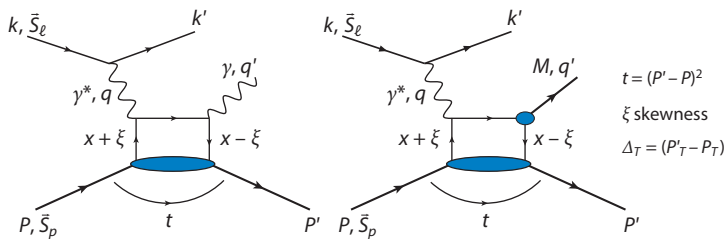

**Figure 3**

Deeply virtual Compton scattering and deeply virtual meson production in exclusive deep-inelastic scattering probe the generalized parton distributions, which can be interpreted as the coordinate (impact parameter) space distributions.

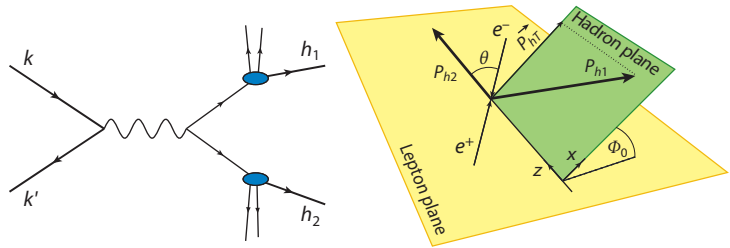

**Figure 4**

Back-to-back dihadron production in $e^+e^-$ annihilations. The hadrons are assumed to be in back-to-back jets and can be applied to the study of transverse momentum–dependent fragmentation functions.

fraction carried by the parton, $k_\perp$ is the transverse momentum, and $r_\perp$ is the coordinate space variable. These functions can be interpreted as the phase-space ($r_\perp, k_\perp$) distribution of the parton in the transverse plane perpendicular to the nucleon momentum direction. Wigner distributions reduce to TMD distributions and GPDs upon integration over certain variables. These TMD distributions and GPDs are experimentally accessible, whereas Wigner distributions, in general, are not.

## 2.2. $e^+e^-$ Annihilations

To constrain the contribution from the fragmentation processes in SIDIS, we have to measure the relevant hadron production processes in $e^+e^-$ annihilations. In particular, the back-to-back dihadron correlation can provide important information on TMD fragmentation functions (56, 76). This process (**Figure 4**) has its own unique phenomena, which were discovered recently by the Belle and BaBar Collaborations (46–48). Two hadrons are produced in the back-to-back correlation kinematics, and a nontrivial fragmentation function called the Collins fragmentation function leads to a novel azimuthal angular asymmetry proportional to $\cos 2\phi_0$.

## 2.3. Nucleon–Nucleon Scattering

The nucleon structure can also be well studied in hadron–hadron collisions. As shown in **Figure 5**, two partons from the incoming nucleons scatter and, in the final state, produce a high-momentum particle, such as a hadron, jet, or vector boson (from virtual photon decays), or a lepton pair. When choosing an appropriate polarization for the incoming nucleons, we can study the associated parton distributions depending on the spin of the nucleon (29, 77).

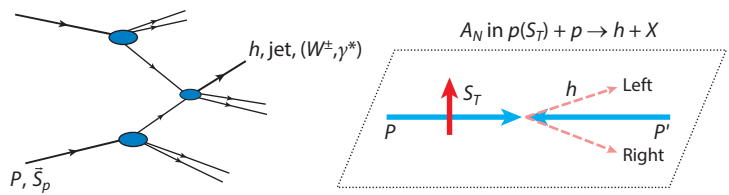

**Figure 5**

Inclusive hadron ($h$), jet, or vector boson production in $pp$ collisions, which depend on the incoming parton distributions of quarks and gluons. Among the spin-dependent observables, the single transverse spin asymmetry (*right*) is closely related to the transverse spin structure of the nucleon. This asymmetry is also called left–right asymmetry.

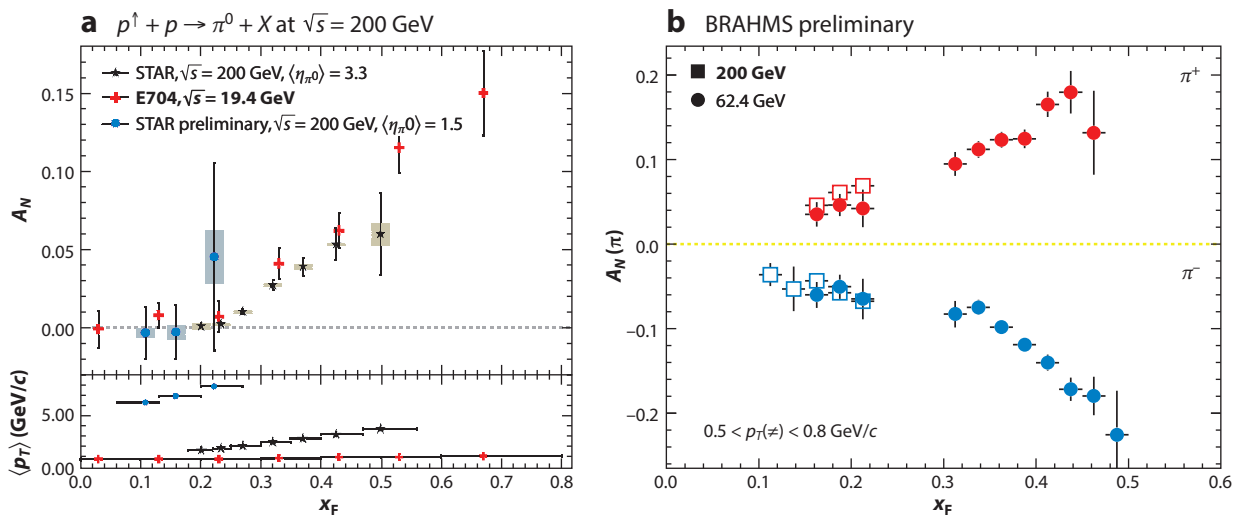

**Figure 6**

Single transverse spin asymmetries in forward hadron [(*a*) neutral pion and (*b*) charged pion] production in *pp* collisions at the Relativistic Heavy Ion Collider as a function of $x_F$. Panel *a* modified from Reference 30. Panel *b* modified from Reference 31.

## 2.4. Transverse Spin Phenomena in High-Energy Experiments

Spin asymmetries in hadronic reactions help unveil the unique correlation of partons inside the nucleon. SSAs are among the early experimental observations of spin asymmetries in high-energy hadron collisions (17, 18), where a beam of transversely polarized protons was scattered off unpolarized nucleons. These experiments (17, 18) showed that the charged pions are produced asymmetrically to the left or right of the plane spanned by the momentum and spin directions of the initial polarized protons:

$$A_N = \frac{d\sigma(S_\perp) - d\sigma(-S_\perp)}{d\sigma(S_\perp) + d\sigma(-S_\perp)}. \qquad 1.$$

Therefore, this asymmetry is also called left–right asymmetry (**Figure 5**). Subsequent observations—in particular, a series of experiments at the Fermilab fixed-target facility (19–22)—have confirmed these early findings. In the past decade, such experiments have been extended to the collider energy region at RHIC at BNL (23–29); **Figure 6** summarizes examples for various collision energies. The results exhibit a general pattern: Sizable asymmetries are measured at forward rapidity and for positive Feynman $x_F > 0.3$, which increases in magnitude with increasing $x_F$. In contrast, for negative $x_F$ and at midrapidity, all asymmetries are consistent with zero.

Meanwhile, in DIS experiments, the HERMES, COMPASS, and JLab Collaborations have observed large SSAs in SIDIS. In SIDIS, $e(k) + p(P) \rightarrow e(k') + h(P_h) + X$ under kinematic conditions (**Figure 2**), where $P_{h\perp} \ll Q$. The differential cross section may be noted as

$$\frac{d\sigma}{dx_B \, dy \, dz_h \, d^2 \vec{P}_{h\perp}} = \frac{4\pi\alpha_{em}^2 s}{Q^4} \Big[ (1 - y + y^2/2)x_B(F_{UU} - \sin(\phi_h - \phi_S)|\vec{S}_\perp| F_{UT}^{\text{Sivers}}) \\ - (1-y)x_B|\vec{S}_\perp| \sin(\phi_h + \phi_S) F_{UT}^{\text{Collins}} \Big], \qquad 2.$$

where $\sigma_0^{(\text{DIS})} = 4\pi\alpha_{em}^2 S_{ep}/Q^4 \times (1 - y + y^2/2)x_B$ with the usual DIS kinematic variables $y$, $x_B$, and $Q^2$; $z_h = P_h \cdot P/q \cdot P$; $P_{h\perp}$ is the transverse momentum of the final-state hadron with respect to the virtual photon direction; and $\phi_S$ and $\phi_h$ are the azimuthal angles of the proton's transverse polarization vector and the transverse momentum vector of the final-state hadron, respectively. The Sivers and Collins contributions noted above are the two most important contributions to SSAs. The associated asymmetries can be obtained by taking different azimuthal angular distributions of the differential cross sections: The Sivers contribution is proportional to $\sin(\phi_h - \phi_S)$, and the Collins contribution is proportional to $\sin(\phi_h + \phi_S)$. The nonzero value of Sivers SSAs observed by the HERMES Collaboration in early 2000, in particular, has stimulated great interest in the hadron spin physics community. Theoretical interpretation of these results can be traced back to the fundamental parton distributions in a transversely polarized nucleon. **Figure 7** shows an example of a comparison of the current experimental data on these two asymmetries with the proton target between the HERMES and COMPASS Collaborations. As shown, large asymmetries have been observed for both terms in the valence region of DIS experiments.

In addition, hadron pair production in $e^+e^-$ annihilation has also shown significant azimuthal asymmetries, as observed by the Belle and BaBar Collaborations (46–48). These experiments focus on back-to-back dihadron productions, $e^+ + e^- \rightarrow h_1 + h_2 + X$, with center-of-mass energy $s = Q^2 = (P_{e^+} + P_{e^-})^2$ and the final-state two hadrons with momenta $P_{h1}$ and $P_{h2}$, respectively, as shown in **Figure 4**. The longitudinal momentum fractions are further identified: $z_{hi} = 2|P_{hi}|/Q$. In this process, a quark–antiquark pair is produced from the $e^+e^-$ annihilations at leading order. This pair then fragments and produces the two final-state hadrons. Collins fragmentation functions lead to $\cos 2\phi$ azimuthal angular asymmetries between these two hadrons (56):

$$\frac{d^5\sigma^{e^+e^-\rightarrow h_1 h_2 + X}}{dz_{h1}dz_{h2}d^2 P_{h\perp}d\cos\theta} = \frac{N_c \pi \alpha_{\rm em}^2}{2 Q^2}[(1 + \cos^2\theta)Z_{uu}^{h_1 h_2} + \sin^2\theta \cos(2\phi_0)Z_{\rm Collins}^{h_1 h_2}], \qquad 3.$$

where $\theta$ is the polar angle between hadron $h_2$ and the beam of $e^+e^-$, $\phi_0$ is defined as the azimuthal angle of hadron $h_1$ relative to that of hadron $h_2$, and $P_{h\perp}$ is the transverse momentum of hadron $h_1$ in this frame (**Figure 4**). In Equation 3, $Z_{uu}$ is derived from two unpolarized quark fragmentation functions, whereas $Z_{\rm Collins}$ is derived from two Collins fragmentation functions. To eliminate false asymmetries, the Belle and BaBar Collaborations consider the ratios of unlike-sign "$U$" $(\pi^+\pi^- + \pi^-\pi^+)$ over like-sign "$L$" $(\pi^+\pi^+ + \pi^-\pi^-)$ or charged "$C$" $(\pi^+\pi^+ + \pi^-\pi^- + \pi^+\pi^- + \pi^-\pi^+)$ pion pairs. For example, one of the asymmetries is measured as $A_0^{UL}$,

$$A_0^{UL}(z_1, z_2, \theta, P_{h\perp}) \equiv \frac{\langle \sin^2\theta\rangle}{\langle 1 + \cos^2\theta\rangle}\left(\frac{Z_{\rm Collins}^U}{Z_{uu}^U} - \frac{Z_{\rm Collins}^L}{Z_{uu}^L}\right), \qquad 4.$$

and similarly for $A_0^{UC}$. **Figure 8** provides examples of experimental data from the Belle and BaBar Collaborations, where the asymmetries are plotted as functions of $z_1$ and $z_2$ (the transverse momentum is integrated out). Recently, the BaBar Collaboration also measured the transverse momentum dependence of the above asymmetries. Future experiments may incorporate more data on these fragmentation function measurements.

# 3. TRANSVERSE SPIN STRUCTURE OF THE NUCLEON

In the following sections, we review the basics of the transverse spin structure of the nucleon, which can be extracted from the experiments discussed in the previous sections.

## 3.1. Parton Distributions in a Transversely Polarized Nucleon

In high-energy hadron collisions, various parton distributions can be probed. During the theoretical development of the transverse spin structure in the early days of hadron physics research,

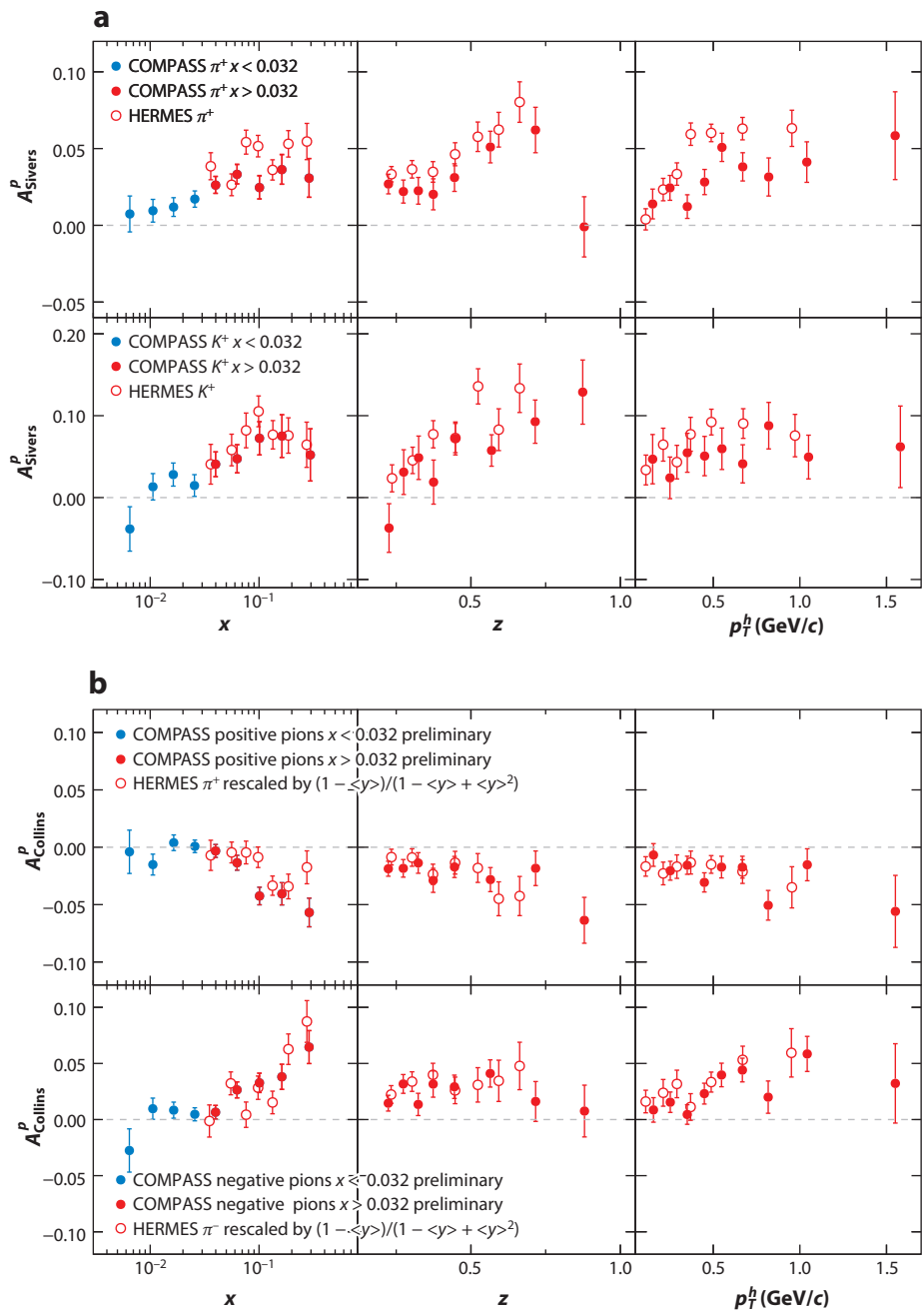

**Figure 7**

Summary of (*a*) Sivers and (*b*) Collins single transverse spin asymmetry measurements in semi-inclusive deep-inelastic scattering on the proton target from the HERMES and COMPASS experiments. Modified from Reference 32.

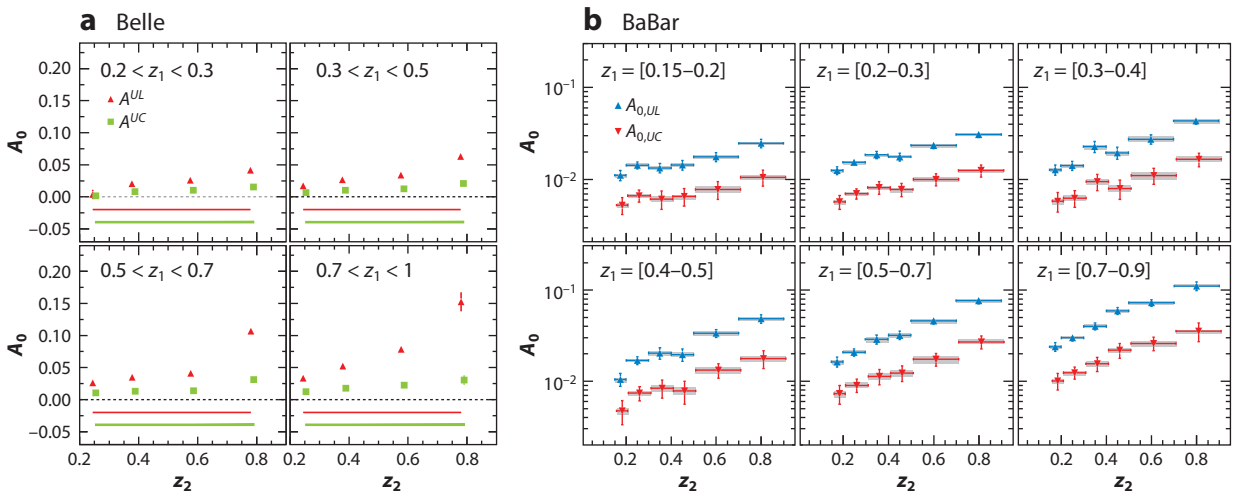

**Figure 8**

Dihadron azimuthal asymmetries in $e^+e^-$ annihilation from the (*a*) Belle (47) and (*b*) BaBar (48) experiments. $A_0^{UL}$ and $A_0^{UC}$ are functions of $z_1$ and $z_2$ for different bins.

two parton distributions were the main focus of discussions: the quark transversity distribution and the Sivers function for quarks and gluons. The quark transversity distribution is one of the three leading-twist-integrated quark distributions (78–80) for the case in which both the nucleon and the quark are transversely polarized. The lowest moment of this distribution provides information on the quark contribution to the nucleon tensor charge. The nucleon tensor charge is one of the fundamental properties of the nucleon and can be computed from lattice gauge theory (81). The quark transversity distribution is difficult to study in normal high-energy scattering processes because of its odd chirality.

The Sivers function was introduced in the late 1980s to explain the large SSAs in hadron production in $pp$ collisions (82, 83). It describes the correlation between the transverse momentum of partons and the polarization vector of the nucleon. By intuitive arguments, the asymmetric distributions of partons in the transversely polarized nucleon lead to a Sivers asymmetry in hadron production in $pp$ collisions. Ideas from both the quark transversity distribution function and the parton Sivers function have been applied to hadron processes to explain the large SSAs observed by experiments (84–89), thus generating strong interest in phenomenology.

In the past two decades, advances in both theory and experiments have enabled progress in achieving a complete picture of nucleon tomography in terms of GPDs and TMD distributions. Within this framework, the transverse spin structure of the nucleon may be fully investigated. In particular, TMD distributions, such as the parton Sivers functions, have been rigorously defined in the hard scattering processes.

In nucleon tomography, both GPDs and TMD distributions provide additional information on parton distributions in the transverse plane. Spin-averaged quark distributions are symmetric in this plane. However, if the nucleon (or the quark) is transversely polarized, the quark distribution will be azimuthally asymmetric in the transverse plane. The TMD quark Sivers function and the GPD E distribution function quantify these asymmetries in the transverse momentum space and in the coordinate space, respectively. **Figure 9** shows these distributions for the $u$ quark (averaged over the longitudinal momentum fraction $x$) in the impact parameter space and the transverse momentum space in a transversely polarized nucleon. The impact parameter–dependent quark

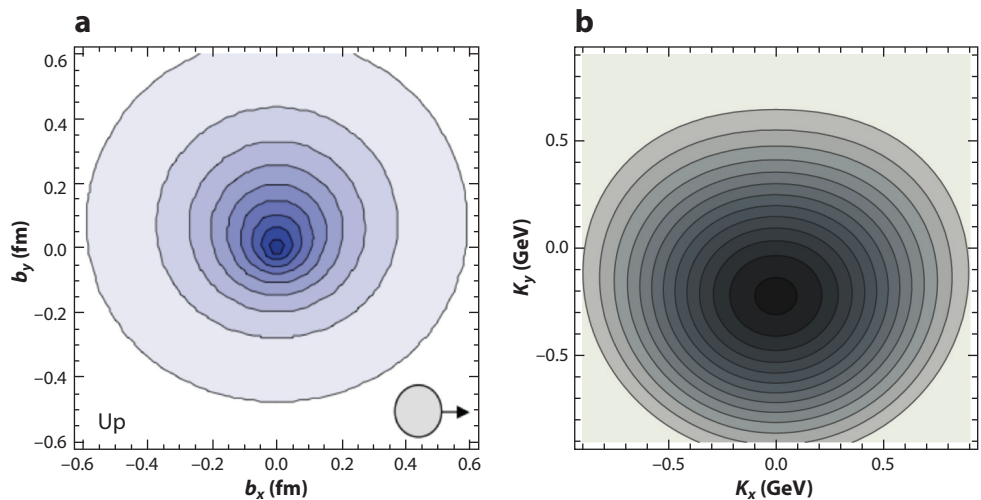

Transverse distortion of the quark distribution when the nucleon is transversely polarized along the $\hat{x}$ direction (*a*) in the coordinate space (i.e., the impact parameter space) (from Reference 90) and (*b*) in the momentum space (from the fit in Reference 92).

distribution comes from the lattice QCD calculations (90), whereas the TMD distribution comes from the fit to the associated SSA in SIDIS hadron production processes (91, 92).

The plots shown in **Figure 9** compose just part of the tomography picture of the nucleon that may be obtained from GPDs and TMD distributions. Owing to the limited experimental constraints, they show only some qualitative features of the three-dimensional picture of the quark in the nucleon. The model dependence of these distributions will be improved by future experiments at RHIC, JLab, and COMPASS. The planned EIC, in particular, will focus on both GPDs and TMD distributions in great detail and will provide a comprehensive tomography of the partons inside the nucleon (65, 93).

The tomography picture of parton distributions has also motivated recent theoretical developments to interpret the nucleon spin sum rule for a transversely polarized nucleon. Asymmetric distributions of quarks and gluons in the transverse plane (shown in **Figure 9***a*) contribute to the nucleon transverse spin in the rest frame (94, 95). In a moving frame, we have to consider the Lorentz-invariant Pauli–Lubanski spin vector, which generalizes the spin operator in the rest frame of the system (96, 97). However, a complete transverse polarization sum rule contains higher-twist contributions, thus becoming much more complicated (97–103).

## 3.2. Hard QCD Processes to Explore Generalized Parton Distributions and Transverse Momentum–Dependent Distributions

In terms of experimental measurements, parton distributions are probed in hard scattering processes where short-distance physics takes place and hard interaction knocks out the parton from the nucleon bound state. An immediate advantage of hard scattering processes is that they guarantee the leading power interpretation of the parton distributions probed in experiments.

GPDs are probed in hard exclusive processes in lepton–nucleon scattering, including deeply virtual Compton scattering and deeply virtual meson production processes. Theoretical developments in the past two decades have laid solid foundations to apply QCD theory to describe these processes and explore the associated GPDs, which have been very well summarized in recent reviews (60–65).

For TMD distributions, theoretical developments are still ongoing, with many recent advances. New results are reviewed in Section 4. Here, we briefly introduce the hard processes that can be applied to studies of TMD distributions.

We begin with some comments on the general application of TMD distributions to high-energy scattering processes. Although the idea of intrinsic transverse momentum of partons was introduced in the 1970s for various hadron processes, a series of papers by Collins and Soper in the 1980s (9, 109, 110) showed that the concept of TMD distributions makes sense only for hard scattering processes, which involve a hard momentum scale in addition to the transverse momentum in the final-state particle production. This immediately restricted use of TMD distributions to the following processes:

1. In SIDIS hadron production, the hard momentum scale is the virtuality of the exchanged photon $q^2 = -Q^2$. In the low transverse momentum region—namely $P_{b\perp} \ll Q$, where $P_{b\perp}$ is the final-state hadron transverse momentum—we can apply TMD quark distributions and fragmentation functions.
2. Drell–Yan lepton pair production in hadron collisions also includes electroweak boson ($W/Z$) and Higgs boson production processes. Here again, the invariant mass of the lepton pair (or mass of the $W/Z$ boson) acts as the hard momentum scale in addition to the transverse momentum of the pair. In the low–transverse momentum region $q_\perp \ll Q$, we can apply TMD distributions.
3. In back-to-back dihadron production in $e^+e^-$ annihilations, when the total energy is the hard momentum scale, the relative transverse momentum between the two hadrons probes the TMD fragmentation functions.

TMD distributions are defined through the following matrix:

$$\mathcal{M}_{\alpha\beta}(x, k_\perp) = \int \frac{d\xi^- d^2\xi_\perp}{(2\pi)^3} e^{-ix P^+ \cdot \xi^- + i\vec{k}_\perp \cdot \vec{\xi}_\perp} \langle PS | \bar{\psi}_\beta(\xi^-, \xi_\perp) \mathcal{L}^\dagger_{\xi^-, \xi_\perp} \mathcal{L}_{0^-, 0_\perp} \psi_\alpha(0) | PS \rangle. \qquad 5.$$

Here, we assume that the proton is moving along the $+\hat{z}$ direction. The light-cone variables $P^\pm$ are defined as $P^\pm = (P^0 \pm P^z)/\sqrt{2}$. Therefore, the proton momentum is dominated by its plus component $P^+$. Compared with the integrated parton distributions, the two quark fields are separated not only by the light-cone distance $\xi^-$ but also by the transverse distance $\xi_\perp$, which is conjugate to the transverse momentum of the quark $k_\perp$. Because of this difference, the gauge links play important roles concerning the universality of TMD distributions (see the discussion in Section 4, below).

The leading-order expansion contains eight quark distributions (10–12):

$$\mathcal{M}_{\alpha\beta} = \frac{1}{2} \Big[ f_1(x, k_\perp) \not{p} + \frac{1}{M} b_1^\perp(x, k_\perp) \sigma^{\mu\nu} k_\mu p_\nu + g_{1L}(x, k_\perp) \lambda \gamma_5 \not{p}$$
$$+ \frac{1}{M} g_{1T}(x, k_\perp) \gamma_5 \not{p} (\vec{k}_\perp \cdot \vec{S}_\perp) + \frac{1}{M} h_{1L} \lambda i \sigma_{\mu\nu} \gamma_5 p^\mu k_\perp^\nu + h_1(x, k_\perp) i \sigma_{\mu\nu} \gamma_5 p^\mu S_\perp^\nu \qquad 6.$$
$$+ \frac{1}{M^2} h_{1T}^\perp(x, k_\perp) i \sigma_{\mu\nu} \gamma_5 p^\mu \left( \vec{k}_\perp \cdot \vec{S}_\perp k_\perp^\nu - \frac{1}{2} \vec{k}_\perp^2 S_\perp^\nu \right) + \frac{1}{M} f_{1T}^\perp(x, k_\perp) \epsilon^{\mu\nu\rho\sigma} \gamma_\mu p_\nu k_\rho S_\sigma \Big].$$

These distributions depend on the polarizations of the quark and the nucleon states, where $M$ is the nucleon mass. The physical interpretation for these quark distributions is shown in **Figure 10**. For an unpolarized nucleon target, an unpolarized quark distribution $f_1(x, k_\perp)$ and a naïve time reversal–odd transversely polarized quark distribution $h_1^\perp(x, k_\perp)$ (the Boer–Mulders

|  |  | **Quark polarization** | | |
|---|---|---|---|---|
|  |  | **Unpolarized** (U) | **Longitudinally polarized** (L) | **Transversely polarized** (T) |
| **Nucleon polarization** | **U** | $f_1 =$ | | $h_1^\perp =$  Boer–Mulder |
|  | **L** | | $g_1 =$  Helicity | $h_{1L}^\perp =$ |
|  | **T** | $f_{1T}^\perp =$  Sivers | $g_{1T}^\perp =$ | $h_{1T} =$  Transversity $h_{1T}^\perp =$ |

Nucleon spin   Quark spin

**Figure 10**

Leading-twist transverse momentum–dependent (TMD) distributions classified according to the polarizations of the quark ($f$, $g$, $h$) and nucleon ($U$, $L$, $T$). Distributions $f_{1T}^\perp$ and $h_1^\perp$ are called naïve time reversal–odd TMD distributions. For gluons, a similar classification of TMD distributions exists (plot from Reference 93).

function) arising from initial/final-state interactions may be introduced. For a longitudinally polarized nucleon, a longitudinally polarized quark distribution $g_{1L}(x, k_\perp)$ and a transversely polarized distribution $h_{1L}^\perp(x, k_\perp)$ are introduced. Finally, for a transversely polarized nucleon, a quark spin-independent distribution $f_{1T}^\perp(x, k_\perp)$ (the Sivers function) arising from initial/final-state interactions and longitudinally polarized quark polarization $g_{1T}(x, k_\perp)$, a symmetrical transversely polarized quark distribution $h_1(x, k_\perp)$, and an asymmetric transversely polarized quark distribution $h_{1T}^\perp(x, k_\perp)$ are introduced.

Of the eight TMD distributions, three are associated with the $k_\perp$-even structure under the exchange $k_\perp \rightarrow -k_\perp$: $f_1(x, k_\perp)$, $g_{1L}(x, k_\perp)$, and $h_1(x, k_\perp)$, which correspond to the unpolarized, longitudinally polarized, and transversity distributions, respectively. These distributions survive after integrating over the transverse momentum. The other five distributions are associated with the $k_\perp$-odd structures and, hence, vanish when $k_\perp$ are integrated for $\mathcal{M}_{\alpha\beta}$. Both the Sivers function $f_{1T}^\perp(x, k_\perp)$ and the Boer–Mulders function $h_1^\perp(x, k_\perp)$ are odd under naïve time reversal transformation (55) and require initial/final-state interactions to be nonzero (49, 51, 52, 54).

In addition, we define the following matrix for a quark fragmentation into a (pseudo)scalar hadron:

$$\mathcal{M}_b(z_b, p_\perp) = \frac{n^-}{z_b} \int \frac{d\xi^+}{2\pi} \frac{d^2\vec{\xi}}{(2\pi)^2} e^{-i(k^- \xi^+ - \vec{k}_\perp \cdot \vec{\xi}_\perp)} \sum_X \frac{1}{3} \sum_a \langle 0| \mathcal{L}_0 \psi_\beta(0) | P_b X \rangle$$
$$\times \langle P_b X | \bar{\psi}_\alpha(\xi^+, \vec{\xi}_\perp) \mathcal{L}_\xi^\dagger | 0 \rangle,$$

where $\mathcal{L}_{\xi, \xi_\perp}$ is the associated gauge link for the fragmentation function. The quark momentum is dominated by its minus component, and the hadrons' momentum is dominated by their minus component. Thus, $k^- = P_b^-/z$ and $\vec{k}_\perp = -\vec{p}_\perp/z$, where $p_\perp$ is the hadron transverse momentum relative to the quark jet axis. We have also introduced a vector $n = (0^+, 1^-, 0_\perp)$. At leading order, we have two fragmentation functions in the expansion:

$$\mathcal{M}_b = \frac{1}{2}\left[\hat{D}(z_b, p_\perp)\,\slashed{n} + \frac{1}{M}H_1^\perp(z_b, p_\perp)\sigma^{\mu\nu}p_{\mu\perp}n_\nu\right]. \qquad 7.$$

The second is a (naïve) time reversal–odd Collins fragmentation function (55). It represents a correlation between the transverse spin of the fragmenting quark and the transverse momentum of the hadron relative to the jet axis in the fragmentation process.

# 4. QCD THEORY DEVELOPMENTS FOR TRANSVERSE MOMENTUM–DEPENDENT DISTRIBUTIONS

In this section, we review recent developments from QCD theory on TMD distributions, including gauge invariance, universality, factorization, and evolution. These are the theoretical foundations for the phenomenological applications discussed in Section 5.

## 4.1. Gauge Invariance and Universality of the Transverse Momentum–Dependent Distributions

A recent breakthrough in TMD physics started with a model calculation of the SSA in the SIDIS process (49), where the final-state interaction leads to a nonzero SSA in the Bjorken limit ($Q^2 \to \infty$). This final-state interaction is factorized into the gauge link in the gauge-invariant TMD quark distribution (51, 52), and the asymmetry comes from the relevant quark Sivers function (82). In terms of the Drell–Yan process, the final-state interaction becomes the initial-state interaction, and the associated SSA changes sign (49, 51).

**Figure 11** shows examples of the final-state interaction contribution in DIS and initial-state interactions in Drell–Yan lepton pair production processes. In the DIS process, the final-state interaction can be summarized as follows (52):

$$\int \frac{d^4k_g}{(2\pi)^4}\bar{u}(k)(-ig\gamma^\alpha T_a)\frac{i(\slashed{k}-\slashed{k}_g)}{(k-k_g)^2+i\epsilon}\cdots\langle n|\psi(0)A_{a\alpha}(k_g)|P\rangle,$$

where $n$ represents the intermediate state, $q$ is the virtual photon's momentum with $q^2 = -Q^2$, $k$ is the outgoing quark jet's momentum with $k^- \sim Q$, $k_g$ is the gluon momentum toward the proton target, and $a$ is the associated color index. In the Bjorken limit, $Q^2 \to \infty$, the leading contribution

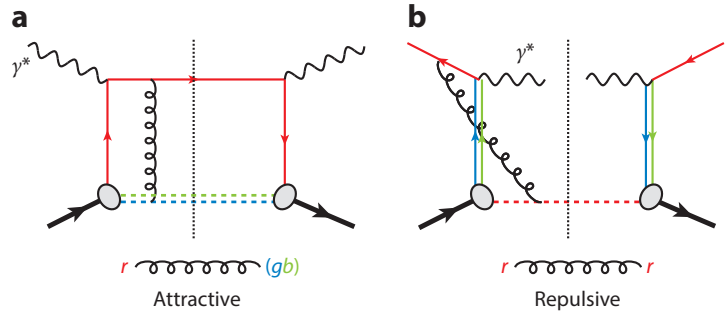

**Figure 11**

Process dependence of the Sivers functions in (*a*) deep-inelastic scattering and (*b*) Drell–Yan processes.

for this diagram comes from the $A^+$ component of the gauge potential, and we obtain

$$(-ig) \int_{-\infty}^{+\infty} dk_g^+ \frac{i}{-k_g^+ + i\epsilon} A^+(k_g) = (-ig) \int_0^\infty d\xi^- A^+(\xi^-), \qquad 8.$$

which is just one order expansion of the gauge link $\mathcal{L}_0$,

$$\mathcal{L}_n(\infty; \xi) \equiv \exp\left(-ig \int_0^\infty d\lambda n \cdot A(\lambda n + \xi)\right), \qquad 9.$$

in the definition of parton distribution functions. The final-state interaction in this process results in a gauge link that goes to $+\infty$. However, in the Drell–Yan lepton pair production process, we have initial-state interactions. The contribution can be summarized as

$$\int \frac{d^4 k_g}{(2\pi)^4} \bar{v}(k)(-ig\gamma^\alpha T_a) \frac{-i(\slashed{k} + \slashed{k}_g)}{(k + k_g)^2 + i\epsilon} \cdots \langle n|\psi(0)A_{a\alpha}(k_g)|P\rangle.$$

When we take the leading-order contributions, we obtain

$$(-ig) \int_{-\infty}^{+\infty} dk_g^+ \frac{i}{-k_g^+ - i\epsilon} A^+(k_g) = (-ig) \int_0^{-\infty} d\xi^- A^+(\xi^-), \qquad 10.$$

which leads to a gauge link that goes to $-\infty$.

Because of the difference in the gauge link directions, a special universality property will result for the so-called naïve time reversal–odd parton distributions between the SIDIS and Drell–Yan processes. For example, for the quark Sivers function (49, 51, 52),

$$f_{1T}^\perp |_{\text{Drell–Yan}} = -f_{1T}^\perp |_{\text{DIS}}, \qquad 11.$$

and the same holds for the Boer–Mulders function $h_1^\perp$. Verifying the predicted nonuniversality of the Sivers functions remains a challenge in strong interaction physics.

By contrast, TMD fragmentation functions are universal among hard processes. For example, the Collins fragmentation function is the same in SIDIS, $e^+e^-$ annihilations, and $pp$ collisions (104–108). This universality is key to extracting the transversity distributions from simultaneously fitting the experimental data from SIDIS and $e^+e^-$ annihilations. The difference between the Sivers distribution function and the Collins fragmentation function is the role of the gauge link. In the Sivers function, the phase generated from the gauge link leads to a nonzero Sivers function, whereas in the Collins function, the phase does not come from the gauge link.

## 4.2. Transverse Momentum–Dependent Factorization and Evolution

The hard processes described in Section 3.2 can be factorized into TMD distributions. For example, the TMD differential cross section in SIDIS can be written in terms of TMD quark distributions and fragmentation functions. Under the full QCD factorization, this intuitive picture does not change significantly, although certain unique features arise when higher-order corrections are considered (109–118). Generic factorization arguments follow those for collinear factorization processes: Higher-order gluon radiations can be factorized into various factors in the factorization formula. For SIDIS, for example, when the gluon radiation is parallel to the incoming nucleon, its contribution belongs to the TMD quark distribution. If the radiated gluon is parallel to the final-state hadron, the contribution is factorized into the TMD fragmentation function. If the radiated gluon is hard, meaning that the momentum is on the order of $Q$, it will contribute to the hard factor.

The TMD factorization is valid in the limit of $Q^2 \to \infty$, and the power corrections of $P_{b\perp}/Q$ have been neglected. This indicates that the TMD factorization applies only in the kinematic region of a transverse momentum much smaller than $Q$, that is, $P_{b\perp} \ll Q$. This provides a powerful

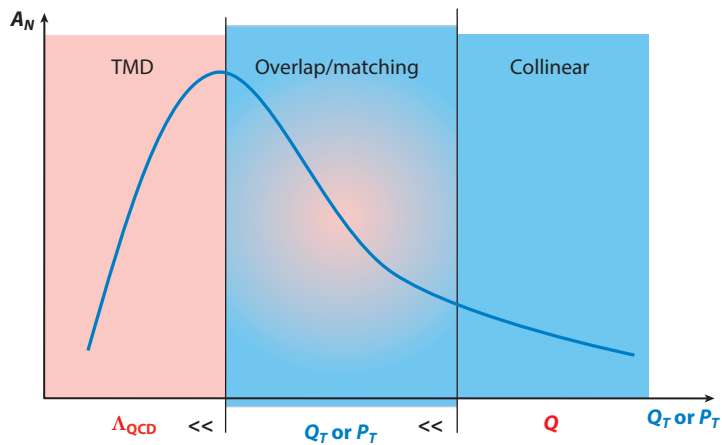

**Figure 12**

Matching the transverse momentum–dependent (TMD) and collinear factorization frameworks for single-spin asymmetries in hard QCD processes, such as semi-inclusive deep-inelastic scattering and Drell–Yan lepton pair production in $pp$ collisions.

description of the transverse spin phenomena in QCD theory. For example, a Sivers-type SSA in SIDIS and Drell–Yan processes depends on the transverse momentum $P_{b\perp}$ (or $q_\perp$ in Drell–Yan) and the hard momentum scale $Q$. Different mechanisms will dominate the physics in different regions of the transverse momentum. At high transverse momentum $P_{b\perp} \sim Q$, the SSA is of higher twist and can be computed from perturbative QCD in the collinear factorization framework. The asymmetry depends on the twist-three quark–gluon–quark correlation function in the nucleon, the so-called Efremov–Teryaev–Qiu–Sterman matrix element (119–123), which can be evaluated from the quark Sivers function $T_F(x, x) = \int d^2k_\perp/M f_{1T}^\perp(x, k_\perp)$. At small $P_{b\perp} \ll Q$, the TMD factorization applies, and the SSA depends on quark Sivers functions. If $P_{b\perp}$ is much larger than $\Lambda_{QCD}$, dependence on the transverse momentum can be computed using QCD perturbation theory. The result obtained from collinear factorization can also be extrapolated into the regime $\Lambda_{QCD} \ll P_{b\perp} \ll Q$, and the result of this extrapolation is identical to that obtained using the TMD approach (124–129). Accordingly, the two mechanisms held to be widely responsible for the observed SSAs are unified. Experimentally, if we can study the transverse momentum dependence of the SSA for a wide range, then we can explore the transition from the perturbative region to the nonperturbative region. This important feature is illustrated in **Figure 12**.

In the factorization arguments discussed above, collinear gluon radiations are factorized into TMD parton distributions. Similar to the integrated parton distribution functions, these gluon radiation contributions can be resummed to all orders by solving the associated evolution equations. In the literature, this resummation is referred to as the TMD or Collins–Soper–Sterman resummation (111). As a result, the factorization simplifies the differential cross section as a convolution of soft factor–subtracted TMD distributions (116), with the factorization scale chosen around the hard momentum scale $\mu_F = Q$. In addition, there is scheme dependence in the TMD definition and factorization, due to the soft factor contribution. The way to subtract the soft factor and regulate the associated light-cone singularity defines the scheme (116). However, after solving the evolution equations, the scheme dependence can be factorized into perturbative calculable coefficients, which can be compared with different schemes. Thus, we obtain a unified picture for TMD phenomenology (130).

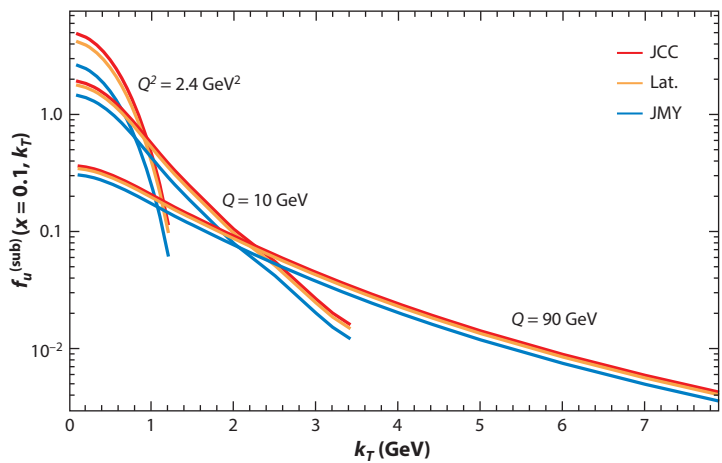

**Figure 13**

The $u$ quark transverse momentum–dependent distributions [$f_u^{(\text{sub})}(x = 0.1, k_{\text{perp}})$] at different scales within chosen schemes corresponding to (*from top to bottom*) JCC, Lat, and JMY (ln $\rho$ = 1) schemes. Abbreviations: JCC, the Collins (2011) scheme (116); JMY, the Ji–Ma–Yuan scheme (112); Lat, lattice. Modified from Reference 130.

Applying the TMD evolution, the unpolarized quark distribution in Equation 6, for example, becomes $f_q(x, k_\perp) \rightarrow f_q^{(\text{sub})}(x, k_\perp, \mu_F = Q)$, which enters into the final factorization formula of Equation 12. This unpolarized quark distribution is the most-studied example. **Figure 13** shows the TMD $u$ quark distribution $f_q^{(\text{sub})}(x = 0.1, k_\perp, \mu_F = Q)$ as a function of the transverse momentum at different scales with three chosen schemes: the Collins (2011) scheme (116) (JCC), the Ji–Ma–Yuan scheme (JMY) (112), and the lattice scheme (Lat) (118). As a general feature, there are broadening effects for TMD distributions at higher scales. These plots also show that the difference between the different schemes becomes less evident at higher scales. Furthermore, the scheme dependence in TMD distributions will be compensated for by the hard factors of each scheme, and the final expressions will be the same when compared with the physical cross sections from experiments (for extensions of the evolution to the Sivers function, see References 131–133). Phenomenological application of the evolution effects to global analysis of the experimental data has recently begun (see Section 5).

## 4.3. Lattice and Model Calculations

As a nonperturbative tool, lattice QCD has been playing an increasingly important role in investigations of the fundamental structure of the nucleon. Relevant to the transverse spin structure, various parton correlation functions have been calculated on lattice, including GPDs and TMD distributions. Early efforts to compute TMD distributions on lattice from References 135 and 136, in particular, have generated great interest within the hadron physics community, although obtaining a comprehensive picture of TMD distributions from first-principles calculations remains a challenge. Another important development in the past couple of years is the proposal to compute parton distributions in Euclidean space (137), including those of TMD distributions (118). Such a development paves a rigorous way to computing the nonperturbative TMD distributions from lattice and directly comparing them to experimental data from, for instance, Drell–Yan lepton pair production. This method overlaps somewhat with that of References 135 and 136. Therefore, a combination of both approaches may be a very promising way to achieve the final goal as stated above.

Meanwhile, model calculations of TMD distributions have also played an important role in theoretical and phenomenological applications. Although these models simplify the complexity of QCD hadron dynamics, they help unravel the nonperturbative aspects of TMD distributions. Several models have been used to calculate TMD distributions: The quark–diquark model is the simplest of all the models. The realistic light-cone wave function model takes into account the three-quark constituent component in the nucleon wave function. The bag model incorporates some aspects of QCD dynamics into its framework of the nucleon bound state. For a detailed account of all these models, see Reference 65 (p. 133).

# 5. PHENOMENOLOGY

Phenomenological applications in TMD physics have always been very fruitful (138–147). In this section, we review some recent phenomenological applications of the theory advances to the TMD observables, with particular focus on the Collins and Sivers SSAs. As described in Section 2.4, the Sivers and Collins effects contribute to the SSAs in SIDIS. The transverse spin–dependent differential cross sections can be written as Equation 2, from which we can calculate the Sivers and Collins SSAs from the associated structure functions:

$$
F_{UU} = H_{UU} \int d^2 \vec{k}_\perp d^2 \vec{p}_\perp f_1(x_B, k_\perp) D(z_b, p_\perp) \delta^{(2)}(z_b \vec{k}_\perp + \vec{p}_\perp - \vec{P}_{b\perp}),
$$

$$
F_{UT}^{\text{Sivers}} = H_{UT}^{\text{Sivers}} \int d^2 \vec{k}_\perp d^2 \vec{p}_\perp \frac{\vec{k}_\perp \cdot \hat{\vec{P}}_{b\perp}}{M} f_{1T}^\perp(x_B, k_\perp) D_q(z_b, p_\perp) \delta^{(2)}(z_b \vec{k}_\perp + \vec{p}_\perp - \vec{P}_{b\perp}), \qquad 12.
$$

$$
F_{UT}^{\text{Collins}} = H_{UT}^{\text{Collins}} \int d^2 \vec{k}_\perp d^2 \vec{p}_\perp \frac{\vec{p}_\perp \cdot \hat{\vec{P}}_{b\perp}}{M_b} h_{1T}(x_B, k_\perp) H_1^\perp(z_b, p_\perp) \delta^{(2)}(z_b \vec{k}_\perp + \vec{p}_\perp - \vec{P}_{b\perp}),
$$

where TMD distributions should be understood as the soft factor–subtracted TMD distributions discussed in Section 4.2 in the full QCD analysis, and $\hat{\vec{P}}_{b\perp}$ is a unit vector of $\vec{P}_{b\perp}$. Hard factors $H_{UU}$ and $H_{UT}$ are the same at one-loop order because of the spin independence of the hard interactions.

## 5.1. Sivers Single-Spin Asymmetries

Early analyses of the experimental data from SIDIS included multiple assumptions: simple Gaussian parameterization for all TMD distributions (or no explicit $k_\perp$ dependence) and, most importantly, no TMD evolution effects (discussed in the previous section) in the leading-order analysis. Nevertheless, these analyses demonstrated a very interesting feature of the experimental constraint on quark Sivers functions: Both $u$ and $d$ quark Sivers functions are sizable but with opposite signs (**Figure 14**).

TMD evolution effects have recently been considered in analyses of Sivers SSAs in SIDIS (92, 149–152). In particular, two recent calculations have shown that the evolution effects are not as strong as previously estimated in the kinematics of the HERMES and COMPASS experiments (92, 152). As such, the quark Sivers functions extracted in References 92 and 152 are similar to those in the leading-order TMD analysis (**Figure 14**). In addition, it has been cross-checked that the implementation of the TMD evolution describes well the unpolarized cross sections in References 92 and 152.

However, only partial evolution effects have been included in the two calculations in References 92 and 152. In Reference 92, a truncated evolution equation was applied, whereas

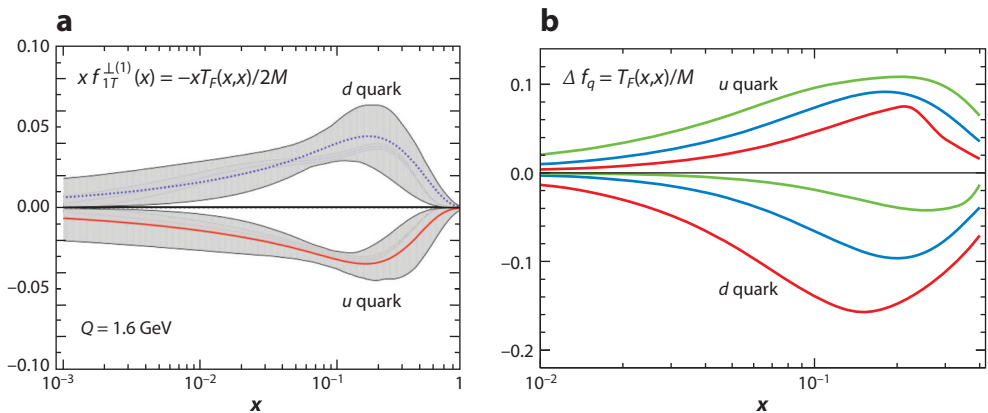

**Figure 14**

Comparison between the leading-order transverse momentum–dependent (TMD) extraction of (*a*) the quark Sivers function from Reference 145 and (*b*) the full QCD extraction with TMD evolution effects taken into account (92).

in Reference 152, the nonperturbative function was modified in the original fit. Further improvements should achieve a consistent full QCD analysis for Sivers SSAs in SIDIS.

## 5.2. Global Analysis of Collins Asymmetries with Full QCD Dynamics

In a series of publications, the Torino group conducted a leading-order analysis for Collins asymmetries in SIDIS and $e^+e^-$ annihilation processes (**Figure 15**) (146, 147). Again in these analyses, simple Gaussian assumptions were made for TMD distribution and fragmentation functions.

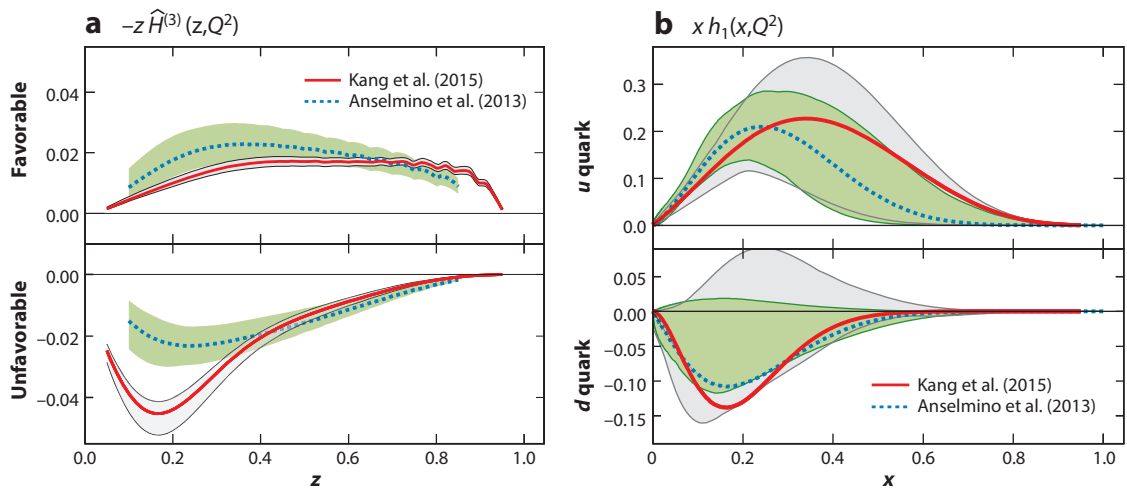

**Figure 15**

Extractions of (*a*) quark transversity distributions and (*b*) Collins fragmentation functions from existing experimental data. The results from References 146 and 147 provide the leading-order transverse momentum–dependent (TMD) analysis, whereas those from Reference 153 are full QCD analyses for which TMD evolution and next-to-leading-order corrections have been taken into account. Modified from Reference 148.

An important step toward a full QCD analysis was made in References 149 and 153, which take into account all relevant theoretical advances in recent years. Three important ingredients are included: (*a*) the perturbative Sudakov form factor at the next-to-leading-logarithmic order, (*b*) the scale evolutions of the integrated quark transversity distribution and the transverse momentum moment of the Collins functions, and (*c*) the hard coefficients in both the unpolarized and transverse spin–dependent differential cross sections at one-loop order. The global fit describes very well the experimental data from both SIDIS and $e^+e^-$ annihilation processes.

The contributions of the transversity distributions extracted from the experiments to the nucleon tensor charge were also estimated in Reference 153:

$$\delta q^{[x_{\min},x_{\max}]}(Q^2) \equiv \int_{x_{\min}}^{x_{\max}} dx \, h_1^q(x, Q^2),$$

$$\delta u^{[0.0065, 0.35]} = +0.30^{+0.08}_{-0.12}, \quad \delta d^{[0.0065, 0.35]} = -0.20^{+0.28}_{-0.11}, \qquad 13.$$

at 90% CL at $Q^2 = 10$ GeV$^2$. This extraction is comparable with previous TMD extractions without evolution (147) and the dihadron method (154, 155). Successful description of the experimental data in the global fit of Collins asymmetries in SIDIS and $e^+e^-$ annihilations in Reference 153 should encourage further phenomenological applications to extend to all other observables associated with TMD distributions. We anticipate more studies along this line in the near future, in particular, for the analyses of Sivers SSAs in SIDIS.

## 5.3. Single Transverse Spin Asymmetries in Hadron Production in *pp* Collisions

The large SSA in hadron production in *pp* collisions initiated the idea of a transverse momentum dependence of quark distributions in polarized protons. However, theoretical understanding of this phenomenon is far from satisfactory. Because the leading-order perturbative calculations vanish for $A_N$ (156), mechanisms beyond the naïve parton model mainly consisting of two proposals have been suggested. At large hadron transverse momentum, the collinear factorization involving twist-three contributions apply, and the asymmetries will be power suppressed $(1/P_{b\perp})$ (123). An alternative approach proposes a generalized parton model that applies the TMD distributions in this process to describe the experimental data (84–89). In the generalized parton model, the SSAs come from the initial-state quark and gluon Sivers functions or the final-state Collins fragmentation function. Phenomenologically, this approach is very attractive. The main issue, however, is the lack of argument to support the factorization for inclusive hadron production in *pp* collisions in terms of TMD distributions (also see the discussion in Section 3.2). Another important issue is the lack of clarity regarding the physical origin of the SSA. As described in Section 4, the SSA comes from the phase generated from initial- or final-state interactions. This phase (for the Sivers function) has not been clearly assessed in this approach.

In the collinear twist-three approach, the twist-three effects can come from the distribution functions of the incoming polarized nucleon (*A*) with momentum $P_A$, from the unpolarized nucleon (*B*) with momentum $P_B$, or from the fragmentation function for the final-state hadron (*h*) with momentum $P_h$. Therefore, the single transverse spin–dependent differential cross section (123, 157) may be noted schematically as follows:

$$d\sigma(S_\perp) = \epsilon_\perp^{\alpha\beta} S_{\perp\alpha} P_{b\beta} \int [dx][dy][dz] \left\{ \phi_{i/A}^{(3)}(x, x') \otimes \phi_{j/B}(y) \otimes D_{b/c}(z) \otimes \mathcal{H}_{ij \to c}^{(A)}(x, x', y, z) \right.$$
$$+ \phi_{i/A}(x) \otimes \phi_{j/B}^{(3)}(y, y') \otimes D_{b/c}(z) \otimes \mathcal{H}_{ij \to c}^{(B)}(x, y, y', z) \qquad 14.$$
$$\left. + \phi_{i/A}(x) \otimes \phi_{j/B}(y) \otimes D_{b/c}^{(3)}(z, z') \otimes \mathcal{H}_{ij \to c}^{(c)}(x, y, z, z') \right\},$$

where $\epsilon_\perp^{\alpha\beta}$ is defined as $\epsilon_\perp^{\alpha\beta} = \epsilon^{\mu\nu\alpha\beta} P_{A\mu} P_{B\nu} / P_A \cdot P_B$ with the convention of $\epsilon^{0123} = 1$. Leading-twist parton distributions are labeled by $\phi_{i/A}(x)$ and $\phi_{j/B}(y)$. The leading-twist parton fragmentation function is represented by $D_{b/c}(z)$, where parton $c$ can be a quark or gluon. In Equation 14, the superscript numeral three represents the twist-three correlations for the distribution functions and the fragmentation functions. For a complete analysis, the above three terms have to be taken into account. The first term in the above equation, the contribution from the twist-three parton correlations $\phi_{i/A}^{(3)}(x, x')$ from the polarized nucleon, has been calculated in References 123 and 157–163, which can be related to quark and gluon Sivers functions. The second term from the twist-three effect in the unpolarized nucleon $\phi_{j/B}^{(3)}(x, x')$ is very small in the forward region of the polarized nucleon, where most experimental data exist (164). The third term is different from the other two because it comes from the twist-three fragmentation functions. This term can be related to the Collins fragmentation function. There have been developments in recent years to compute the third term (165–167), which might be the dominant contribution. A complete understanding of the long-standing puzzle of $A_N$ in $pp \rightarrow b + x$ will depend on a comprehensive study of all possible contributions. To do that, more precise parameterizations of the Sivers and Collins functions are needed.

## 6. FUTURE EXPERIMENTS AND THEIR IMPACT

As evident from the discussion throughout this review, transverse spin physics has captured significant interest within the field of hadron physics. Abundant experimental observations of SSAs in different high-energy reactions involving hadrons have stimulated broad theoretical progress over the past decade, which in turn has led to predictions for new observables to complete and further deepen the understanding of transverse spin phenomena. As a result, there are several proposals for future experiments, including the measurement of SSAs in Drell–Yan-like processes, high-precision SIDIS measurements, and fragmentation function studies in $e^+e^-$ annihilations.

### 6.1. Sign Change of Sivers Asymmetries

Several experimental collaborations have proposed complementary ways of measuring Sivers asymmetries in Drell–Yan lepton pair production in hadron collisions to confirm the predicted sign change between Drell–Yan and SIDIS. In this section, we briefly discuss three experimental proposals that are expected to be carried out in the near future: the COMPASS proposal at CERN, the STAR measurement from RHIC at BNL, and two new SeaQuest proposals at Fermilab.

The COMPASS Collaboration started data collection in June 2015, scattering a secondary negative pion beam from the SPS with an energy of 190 GeV from a polarized hydrogen ($NH_3$) target. Experimental details and the kinematic coverage for this measurement can be found in Reference 168. A unique feature of the COMPASS Drell–Yan measurement lies in the $\bar{u}$ and $d$ valence structure of the $\pi^-$ that favors Drell–Yan reactions annihilating $u$ quarks from the polarized proton target with $\bar{u}$ quarks from the beam pion. Importantly, the COMPASS acceptance for the $u$ quark Sivers function in the target is highest for $0.1 < x_u < 0.3$. This is the region for which the $u$ quark Sivers function has been constrained through the past SIDIS measurements of the HERMES and COMPASS Collaborations. Therefore, theory uncertainties in the comparison of the SIDIS and Drell–Yan measurements are expected to be relatively small. When TMD evolutions are taken into account, the predicted asymmetries from References 92 and 152 are sizable, and a significant measurement will be possible given the projected experimental sensitivities. COMPASS is expected to dedicate its data collection campaigns in 2015 and 2018 to polarized Drell–Yan physics.

Two future polarized Drell–Yan experiments have been proposed for the SeaQuest spectrometer at Fermilab (169, 170): the first with a polarized target and the second with a polarized beam with an incoming proton beam energy of 120 GeV. Different from the Drell–Yan experiments at COMPASS, the Fermilab proposals focus on $pp$ scattering. Therefore, the quark flavor structure will be different from that in COMPASS. For the SSA measurement with a polarized beam, the experiment will probe the valence region of the polarized proton, $x_{beam} > 0.3$. This will extend the kinematic region probed by the past SIDIS experiments and the current Drell–Yan measurement in COMPASS.

By contrast, for measurements with polarized target in SeaQuest, $x_{beam}$ will still be in the valence region, and with the beam quark distribution dominated by valence $u$ quarks, the observed Sivers SSA will be sensitive to the $\bar{u}$ quark Sivers function in the proton target. Observation of nonzero asymmetries would signal nonzero quark Sivers distributions for the sea.

Drell–Yan lepton pair production at RHIC has been considered for many years. Given experimental limitations, it has been realized only recently that a measurement of Sivers SSAs in $W^{\pm}$ boson production in polarized $pp$ collisions at $\sqrt{s} = 510$ GeV will be the better approach (29). This process has two important features. First is the unique flavor information: The $W^{+}$ asymmetry probes the $u$ quark and $\bar{d}$ quark Sivers functions in the polarized proton, whereas the $W^{-}$ asymmetry is sensitive to the $d$ quark and $\bar{u}$ quark Sivers functions. Therefore, the separate measurement of SSAs for $W^{\pm}$ bosons will provide unique flavor information for the quark Sivers functions of the proton. Second, because of the large $W$ boson mass, Sivers asymmetries for $W$ bosons will provide a unique opportunity for testing the theoretical understanding of TMD evolution as a function of $Q^2$. SSA measurements for $W$ bosons will be compared with SSAs in lepton pair Drell–Yan at much lower $Q^2$. In addition to SSA measurements for $W$ and $Z$ bosons and Drell–Yan lepton pair production, other channels at RHIC are available for studying the process dependence of the Sivers asymmetries. These include direct photon production and jet production. For the STAR experiment, a dedicated data-taking campaign focusing on $W$ boson SSAs aiming for high luminosity has been tentatively scheduled for 2017.

In summary, the experiments at CERN, RHIC, and Fermilab are complementary. In combination, they will make it possible first to test for the important sign change of the Sivers function between SIDIS and Drell–Yan processes and then to map the quark Sivers distributions as a function of the kinematics and flavor. The greatest experimental challenges seem to lie in achieving the substantial integrated luminosities required with polarized beams and targets for Drell–Yan physics.

## 6.2. Semi-Inclusive Deep-Inelastic Scattering with CEBAF at 12 GeV and a Future Electron–Ion Collider

Measurements constraining GPDs and TMDs are among the central scientific goals of the upgrade of CEBAF from 6 to 12 GeV at JLab (71). Starting in 2016, all transverse spin–dependent TMD distributions will be extensively studied with a broad suite of experiments in combination with polarized proton, deuteron, and helium-3 targets using the electron beam at CEBAF upgraded to a beam energy of 12 GeV with its extraordinary beam intensity and the resulting high statistical precision and large kinematic reach. Precision measurements of SSAs in semi-inclusive hadron production with multidimensional binning in momentum fraction $x$, momentum transfer $Q^2$, final hadron momentum fraction $z_h$, and transverse momentum $P_{h\perp}$ are necessary for a model-independent extraction of TMD distributions.

It appears that the only significant remaining experimental limitation for experiments at the upgraded CEBAF will be the limited range in $Q^2$ for the fixed-target experiments. Only at the

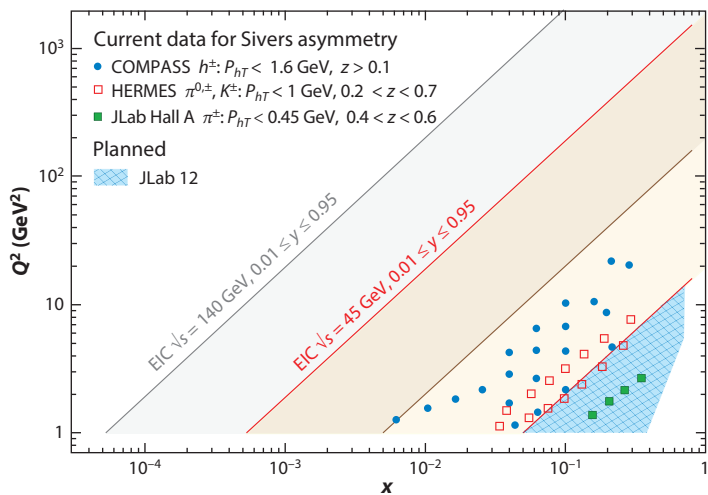

**Figure 16**

Comparison of the kinematic coverage in $x$ and $Q^2$ between existing semi-inclusive deep-inelastic scattering experiments and the planned JLab 12-GeV upgrade and an electron–ion collider (EIC). Modified from Reference 93.

proposed future EIC, which will involve collisions of high-intensity electron beams on high-intensity polarized proton beams at very high $Q^2$, will it become possible to overcome this common limitation of all fixed-target experiments studying nucleon structure. In combination with a comprehensive and hermetic detector, the EIC's ability to provide measurements of SSAs over a wide kinematic range with high luminosity will make it possible to determine the transverse spin structure of the proton with unprecedented precision (93). In parallel to the active TMD and GPD physics program at JLab, the EIC will require a decade of efforts in accelerator and detector development and construction to be able to provide the best possible experimental facility for the systematic and comprehensive study of TMD and GPD physics.

**Figure 16** compares the $x$–$Q^2$ coverage of HERMES, COMPASS, and CEBAF with 12-GeV beam energy with the kinematic coverage of the EIC. The wide kinematic coverage puts the EIC in the unique position of accessing the valence region at much larger $Q^2$ than is possible in current and near-future experiments. As such, QCD dynamics associated with the transverse momentum dependence in hard processes may also be rigorously studied. The planned EIC will also be able to access low-$x$ values reaching down to less than $10^{-4}$, where sea quarks and gluons are abundant and may be studied in detail. The projected high luminosity will also allow for a fully differential analysis over almost the entire kinematic phase space in $x_B$, $Q^2$, $z_b$, and $P_{b\text{perp}}$. This will be critical input for phenomenological analyses.

## 6.3. Fragmentation Function Measurements at $e^+e^-$ Colliders

Data on TMD hadron production in $e^+e^-$ annihilation have provided additional information on transverse spin–dependent quark fragmentation functions compared to the measurements of SSAs carried out in SIDIS and polarized $pp$ collisions. Collins azimuthal asymmetries measured in SIDIS and $pp$ collisions depend on quark transversity distributions in the initial state and spin-dependent fragmentation functions in the final state. By contrast, Collins asymmetries measured in $e^+e^-$ annihilation depend only on the final-state Collins fragmentation function. It follows that only combined global analyses of Collins azimuthal asymmetries measured in SIDIS and $e^+e^-$ annihilation will succeed in constraining quark transversity distributions. The Belle and BaBar measurements of Collins asymmetries have made it possible to determine both quark transversity distributions and Collins fragmentation functions.

In the future, more precise measurements of spin-dependent fragmentation functions will become possible through the luminosity upgrade of the *B* factory at KEK. The luminosity of KEK-B may increase by as much as a factor of 50. Following the improved particle identification detectors of the new Belle II detector, it will be possible to measure Collins asymmetries with high statistical precision for identified hadrons and hadron pairs finely binned in fractional hadron momentum, *z*, and transverse hadron momentum, $p_T$. Further, it will be possible to carry out precision measurements of Collins asymmetries for $\rho$ mesons and $\Lambda$ baryons. Finally, in combination with the improved displaced vertex detection capability of the upgraded Belle II detector, it will be possible to better constrain the presently large systematic uncertainties related to contributions from charm quark production.

Recently, the Beijing electron–positron collider has started to measure the Collins asymmetries at lower center-of-mass energy. The measurements at lower $Q^2$ will make it possible to study the evolution of the Collins fragmentation function. In this context, a possible upgrade of the Beijing collider to a super tau-charm factory with a tunable center-of-mass energy from 2 to 7 GeV will be of significant interest. Precision measurements of Collins asymmetries from Belle II and the super tau-charm factory in Beijing will provide important and complementary information needed for the interpretation of results from future SIDIS experiments at JLab and at a future EIC.

## 7. SUMMARY

In this review, we have highlighted the theoretical and experimental advances made in the investigation of nucleon transverse spin structure. We have focused on the study of Sivers and Collins asymmetries in hard scattering processes, both of which have a clear physical interpretation (170a). The unambiguous extraction of the relevant nucleon parton distributions requires an understanding of the associated QCD dynamics. In particular, because a simultaneous and consistent analysis of data from experiments using different processes at different energy scales is needed, a firm understanding of the evolution of TMD distribution and fragmentation functions is needed. Significant theoretical activities were directed toward gaining an understanding of TMD evolution in recent years. For TMD distribution functions and fragmentation functions, the theoretical advances have made it possible to demonstrate that we have arrived at a unified and consistent framework to understand transverse spin asymmetries observed in SIDIS, Drell–Yan lepton pair production in *pp* collisions, and hadron pair production in $e^+e^-$ annihilations.

Phenomenological application of full QCD in the global analysis of Collins azimuthal asymmetries in SIDIS and $e^+e^-$ annihilations also have been reviewed. In the future, we expect more applications to other transverse spin phenomena, in particular, for the Sivers SSAs in SIDIS and Drell–Yan processes, which are currently being met with significant experimental interest.

As emphasized in this review, transverse spin and TMD physics are undergoing rapid development, with strong theoretical and experimental research efforts worldwide. Owing to space constraints, several interesting new results and ideas have not been discussed in detail in this review. Among them, we mention three important topics. First, in the past few years, progress has been made in extending the TMD framework to the small-*x* region, where gluon saturation at small *x* can be related to TMD distributions at small *x* (see, for example, Reference 171). The so-called $k_t$ factorization at low *x* and high gluon densities is similar to the TMD factorization discussed in this review. In this sense, we have identified a common tool set used by both the spin physics and small-*x* physics communities. Second, quark transversity distributions can be studied through dihadron fragmentation processes in SIDIS, $e^+e^-$ annihilation, and $pp$ collision. There have been experimental measurements of these asymmetries, which have provided constraints on the quark transversity distributions (for further discussion, see References 154 and 155 and references therein). Third, though briefly mentioned in Section 4, the model calculations for TMD distributions should be described in more detail. An important aspect of the TMD physics, these calculations have played a critical role in supporting theoretical and phenomenological developments.

In the near future, we expect rapid progress in transverse spin physics to continue. Several Drell–Yan experiments are under way to measure SSAs with pion or proton beams off polarized proton targets; RHIC experiments at BNL will continue studying the long-standing large $A_N$ puzzle using multiple experimental channels; the CEBAF 12-GeV upgrade at JLab will lead to a strong suite of experiments measuring TMD distributions and GPDs; and the planned EIC will provide the ultimate laboratory for nucleon tomography, including its spin, spin-orbital correlation, and various TMD distributions and GPDs. Significant theoretical efforts have been made successfully to introduce TMD theory into the phenomenological applications of experimental observations. Finally, recent advances in lattice QCD will help our understanding of nucleon structure, including its transverse spin structure, from first-principles computations. We end by recalling the history of surprising observations and developments in the field of transverse spin physics. We certainly are looking forward to the future challenges and surprises in the field.

## DISCLOSURE STATEMENT

The authors are not aware of any affiliations, memberships, funding, or financial holdings that might be perceived as affecting the objectivity of this review.

## ACKNOWLEDGMENTS

We are grateful to our colleagues in both theory and experiments for their collaborations, discussions, and comments. This material is based on work supported by the US Department of Energy, Office of Science, Office of Nuclear Physics, under contract number DE-AC02-05CH11231.